\documentclass[prb,twocolumn,byrevtex,floatfix,superscriptaddress]{revtex4}

\usepackage{gensymb}
\usepackage{amsmath}    
\usepackage{amsfonts}
\usepackage{graphicx}   
\usepackage{verbatim}   
\usepackage{color}      
\usepackage{subfigure}  
\renewcommand{\thefootnote}{\arabic{footnote}}
\usepackage{hyperref}   
\setcitestyle{super}

\begin{document}

\title{\boldmath Magnetic avalanche of non-oxide conductive domain walls \unboldmath}
\author{S. Ghara}
\affiliation{Experimental Physics V, Center for Electronic Correlations and Magnetism, University of Augsburg, 86159 Augsburg, Germany}
\author{K. Geirhos}
\affiliation{Experimental Physics V, Center for Electronic Correlations and Magnetism, University of Augsburg, 86159 Augsburg, Germany}
\author{L. Kuerten}
\affiliation{Department of Materials, ETH Zurich, 8093 Zurich, Switzerland}
\author{P. Lunkenheimer}
\affiliation{Experimental Physics V, Center for Electronic Correlations and Magnetism, University of Augsburg, 86159 Augsburg, Germany}
\author{V. Tsurkan}
\affiliation{Experimental Physics V, Center for Electronic Correlations and Magnetism, University of Augsburg, 86159 Augsburg, Germany}
\affiliation{Institute of Applied Physics, Moldova}
\author{M. Fiebig}
\affiliation{Department of Materials, ETH Zurich, 8093 Zurich, Switzerland}
\author{I. K\'ezsm\'arki\footnote{email: istvan.kezsmarki@physik.uni-augsburg.de}\let\thefootnote\relax\footnote{K.G. and L.K. contributed equally.}}
\affiliation{Experimental Physics V, Center for Electronic Correlations and Magnetism, University of Augsburg, 86159 Augsburg, Germany}


\begin{abstract}
\textbf{Atomically sharp domain walls (DWs) in ferroelectrics are considered as an ideal platform to realize easy-to-reconfigure nanoelectronic building blocks, created, manipulated and erased by external fields. However, conductive DWs have been exclusively observed in oxides, where DW mobility and conductivity is largely influenced by stoichiometry and defects. In contrast, we here report on conductive DWs in the non-oxide ferroelectric GaV$_4$S$_8$, where charge carriers are provided intrinsically by multivalent V$_4$ molecular clusters. We show that this new mechanism gives rise to DWs composed of nanoscale stripes with alternating electron and hole conduction, unimaginable in oxides. By exerting magnetic control on these segments we promote the mobile and effectively 2D DWs into dominating the 3D conductance, triggering abrupt conductance changes as large as eight orders of magnitude. The flexible valency, as origin of these novel hybrid DWs with giant conductivity, demonstrates that non-oxide ferroelectrics can be the source of novel phenomena beyond the realm of oxide electronics.}
\end{abstract}

\maketitle
In order to minimize the energy of a ferroic material, energetically equivalent domain states can form with different orientations of the ferroic order parameter. The number of domain states is dictated by the symmetry reduction upon the para-to-ferroic transition and the population of the domains can be controlled by the conjugate field~\cite{catalan2012, erhart2004}. DWs are geometrically confined regions between adjacent domains, which can permit the formation of spatially confined states with novel functionalities~\cite{aird1998,salje2010}, such as the ferromagnetic and ferroelectric DW states found in antiferromagnetic TbMnO$_3$ and non-polar $A$TiO$_3$ ($A$ = Sr and Ca), respectively~\cite{farokhipoor2014,van2012,zubko2007}.

An intriguing aspect of DWs in ferroelectric insulators is their electrical conductivity, originating from the discontinuity of the normal component of the  polarization across DWs ~\cite{vul1973,gureev2011,eliseev2011,meier2015,sluka2016}. Such conducting DWs have been observed in several materials, like BaTiO$_3$~\cite{sluka2013}, LiNbO$_3$~\cite{schroder2012,werner2017l}, $h$-$R$MnO$_3$ ($R$ = Er and Ho)~\cite{meier2012,wu2012} and BiFeO$_3$~\cite{seidel2009,farokhipoor2011}.

All ferroelectrics in which conductive DWs have so far been reported are oxides, where domain-wall conductivity  usually requires a specific strain configuration of the crystal, an improper character of the ferroelectricity or other unusual properties which render the walls immobile and thus curtail their usefulness and flexibility.  Additionally, oxide materials are prone to defects, significantly hampering the utility of their conducting  DWs.

In this study, we report on highly conductive DWs in a non-oxide ferroelectric, GaV$_4$S$_8$, investigated by macroscopic transport studies as well as by piezoresponse force microscopy (PFM) and conductive atomic force microscopy (c-AFM). This compound belongs to the lacunar spinel family, where the V$_4$ molecular clusters have flexibility to accommodate different number of 3$d$ electrons~\cite{johrendt1998,pocha2000,nakamura2005,muller2006,singh2014,janod2015,reschke2020}, and thus, can provide charged DWs intrinsically with mobile carriers, either conduction electrons or holes, without requiring defects and off-stoichiometry. Hence, this multiferroic material possesses several ingredients that are crucial to realize ideal DW functionalities and are not found to coexist in any oxide material.  For example, these effectively 2D walls govern the overall conductivity of the material even though they constitute only a minor fraction of the 3D bulk volume. Furthermore, a `digital' alternation of electron- and hole-conduction channels occurs on the nanoscale within the conductive DWs of GaV$_4$S$_8$, unprecedented in oxide materials. Finally, this compound is a proper ferroelectric in which conductive DWs emerge in the unstrained crystal spontaneously,  such that they retain their flexibility, facilitating their electric and magnetic control. We demonstrate that the in-situ magnetic erase of DWs, which takes place through an avalanche-like DW expulsion process, leads to a resistive switching and changes the conductance of the material by up to eight orders of magnitude.

GaV$_4$S$_8$ undergoes a  transition from a cubic to a polar rhombohedral phase at $T_{JT}$ $\simeq$ 45\,K, due to a cooperative Jahn-Teller effect lifting the degeneracy of the V$_4$ cluster orbitals~\cite{pocha2000,nakamura2005}. Due to the non-centrosymmetric nature of its cubic paraelectric state and the type of crystal-symmetry lowering ($\bar{4}3m \to 3m$), its ferroelectric state hosts  four polar domain states (P$_1$-P$_4$) with polarization vectors along the four $\langle$111$\rangle$-type cubic body diagonals. Due to the lack of $\pm P$ domains, all the four polarizations span 109$^{\circ}$ with each other, as shown in Figs.~1a and 1b. Lamellar domain patterns, formed of two alternating domains and separated by $\{100\}$-type mechanically and electrically compatible 109$^{\circ}$ DWs, were observed by PFM~\cite{butykai2017,neuber2018}.


\begin{figure*}[t!]
\includegraphics[width=6.9in]{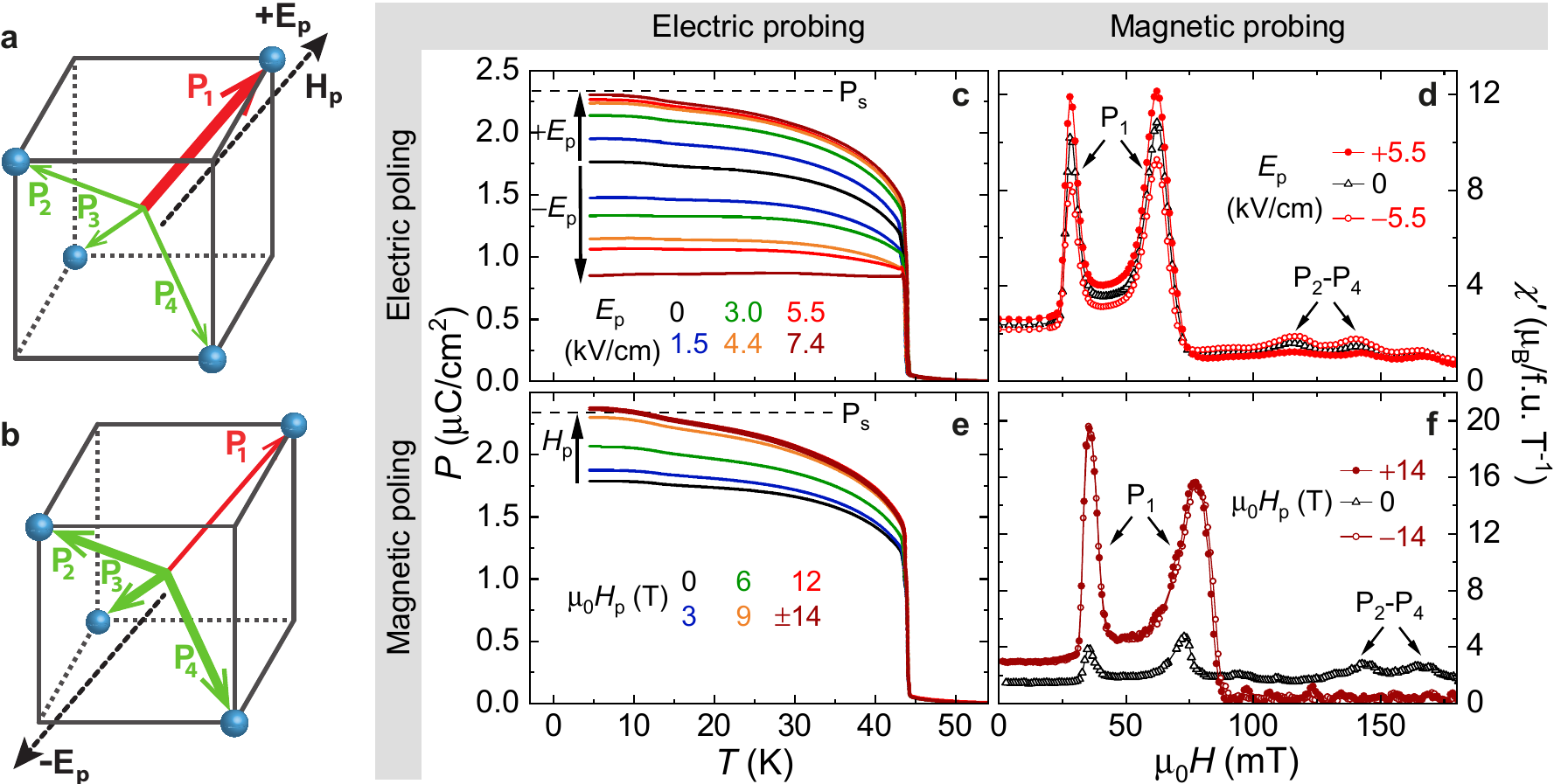}
\caption{\textbf{$\mid$ Electric and magnetic control/readout of multiferroic domain population in GaV$_4$S$_8$.} \textbf{a \& b,}~Schematics about the domain selection process by electric ($E_p$) and magnetic ($H_p$) poling fields applied along the [111] axis. Polarizations of the four domain states, P$_1$-P$_4$, are indicated with respect to the undistorted V$_4$ tetrahedron of the cubic phase. Thick/thin arrows correspond to domains favoured/unfavoured by the poling. \textbf{c/d,}~Temperature-dependent polarization/field-dependent susceptibility at 11.5\,K and their variation with $E_p$. \textbf{e/f,}~Temperature-dependent polarization/field-dependent susceptibility at 11.5\,K and their variation with $H_p$. In all cases polarization along the [111] axis and susceptibility for $H$ $\parallel$[111] were recorded. \label{Fig1}}
\end{figure*}

\vspace{0.4cm}
Within the polar phase, GaV$_4$S$_8$ undergoes a magnetic ordering transition at $T_C$ = 13\,K to a cycloidal (Cyc) state and subsequently to a ferromagnetic (FM) state below 6\,K~\cite{yadav2008,nakamura2009,kezsmarki2015}. By magnetic fields in the sub-Tesla range, the Cyc state is turned to a N\'eel-type skyrmion lattice (SkL) and then to the FM state. The critical fields strongly depend on the orientation of the field with respect to the polar axis, being also the magnetic easy axis~\cite{kezsmarki2015}. In addition to the polarization induced by the rhombohedral distortion, all the three magnetic phases exhibit sizable magnetoelectric polarizations~\cite{ruff2015,widmann2017}.

\vspace{0.2cm}
\textbf{Results and discussion}

\begin{figure*}[t!]
\includegraphics[width=6.9in]{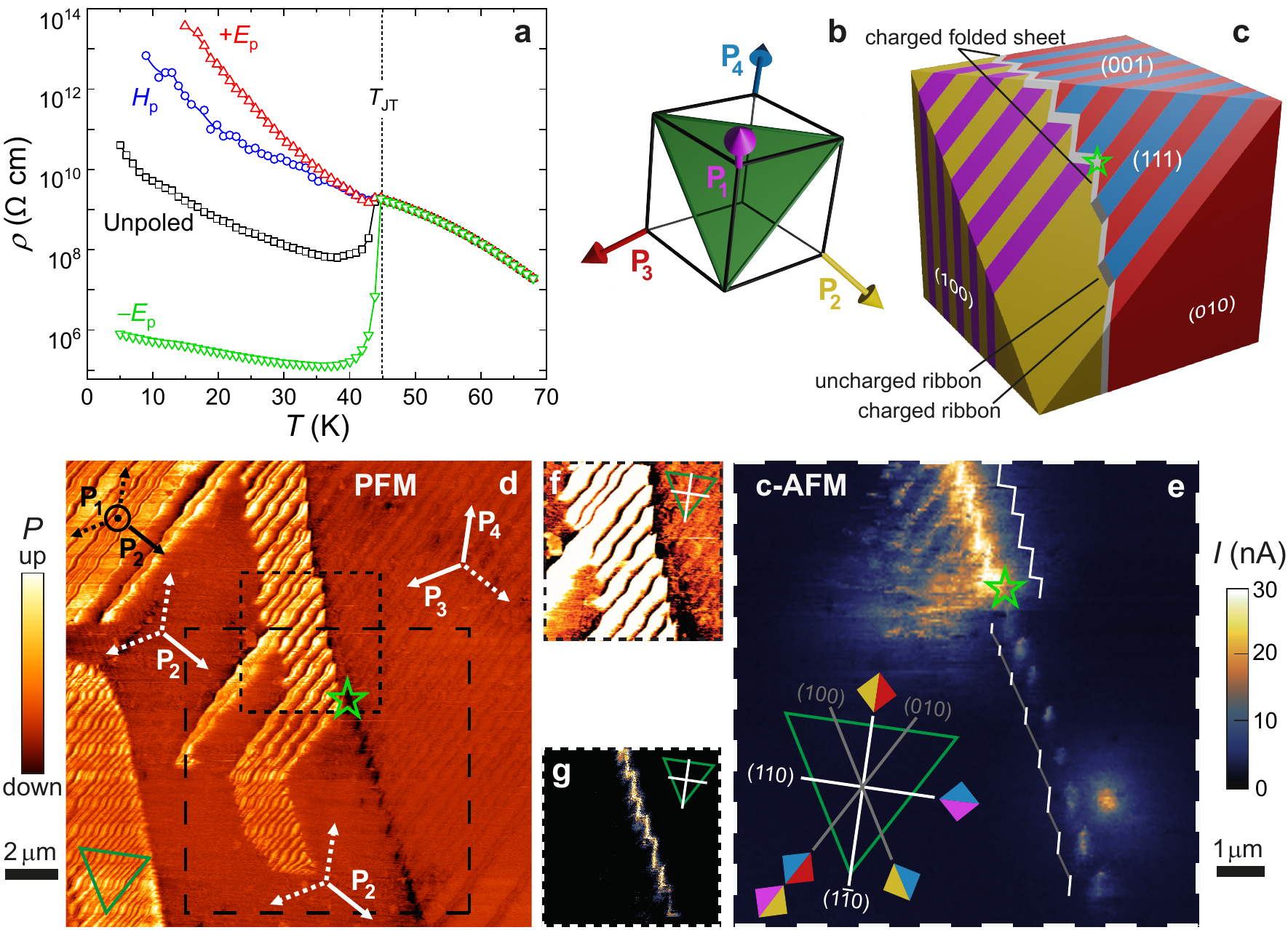}
\caption{\textbf{$\mid$ Macro- and microscopic signatures of conductive DWs in GaV$_4$S$_8$.} \textbf{a,} Dependence of the resistivity on electric ($E_p$=$\pm$5.5\,kV/cm) and magnetic ($\mu_0H_p$=14\,T) poling below $T_{JT}$, implying the presence of conducting DWs. \textbf{b,} Orientation of the polarization in the four polar domains, P$_1$-P$_4$, with respect to the (111) plane (green triangle), which was imaged by PFM and c-AFM. \textbf{c,} Schematic representation of the domain pattern observed in panels \textbf{d-g}. The crystal orientation and the colouring of the domains are according to panel \textbf{b}. White and grey DW segments indicate charged (TT or HH) and uncharged HT junctions, respectively. \textbf{d,} PFM image recorded at 15\,K on the (111) surface of an unpoled sample. 
The polarization directions for domains present/absent in the local domain patterns are indicated by solid/dashed arrows with labeling and orientation according to panel \textbf{b}. \textbf{e,} The c-AFM image, recorded over the region with the thick dashed frame in panel \textbf{d}, confirms the conducting nature of the ribbon- and folded sheet-like DWs, where the folded sheet is a hybrid of alternating HH and TT nanoscale conducting segments, located below and above the green star, respectively. The schematic inset indicates the orientations and assignments of the DWs observed in panels \textbf{d} \& \textbf{e} as well as in the zoomed-in colour-thresholded PFM and c-AFM images of panels \textbf{f} \& \textbf{g}, both corresponding to the region with the thin dashed frame in panel \textbf{d}. Orientation and colour coding follows those in panel \textbf{b} \& \textbf{c}. \label{fig2}}
\end{figure*}

\textbf{Domain wall control and imaging.} Figure 1 shows how the multiferroic nature of GaV$_4$S$_8$ is exploited to control the population of domains by either electric ($E_p$) or magnetic ($H_p$) poling fields and to probe their volume fractions via either electric or magnetic properties. As illustrated in Figs.~1a and 1b, $+E_p$ applied along the [111] axis favours the P$_1$ domain with polarization along the field direction, resulting in the P$_1$ mono-domain state. However, when applying $-E_p$, the material acts as a `poling-diode' due to the lack of inversion domains: Upon the suppression of the P$_1$ domain state, a P$_2$-P$_3$-P$_4$ multi-domain state is created. Figures 1c and 1d, respectively, display the temperature-dependent electric polarization along the [111] axis and the field-dependent magnetic susceptibility at 11.5\,K for magnetic fields parallel to the [111] axis, both measured after different poling runs with $E_p$ $\parallel$[111] (for details of the poling protocol see the Methods part). The unpoled state is dominated by the P$_1$ domain, as reflected by the large positive value of the polarization ($\sim$1.75 $\mu$C/cm$^2$). This domain imbalance, likely arising from internal strains within the crystal, is further supported by the larger magnitudes of two low-field peaks at $\sim$30\,mT and $\sim$70\,mT, as compared to the two high-field anomalies at $\sim$120\,mT and $\sim$140\,mT. The former two are associated with the subsequent Cyc$\to$SkL$\to$FM transitions within the P$_1$ domain with magnetic easy axis parallel to the magnetic field, while the latter two indicate the same transitions taking place simultaneously in the other three domains with easy axes oblique to the field~\cite{kezsmarki2015}. Poling with +$E_p$ enhances the polarization, which saturates at $P_s$$\approx$2.3\,$\mu$C/cm$^2$ for $E_p$$>$+4.4\,kV/cm, resulting in the P$_1$ mono-domain state. In contrast, $-E_p$ decreases the measured polarization, indicating an enhanced volume fraction of the P$_2$, P$_3$, P$_4$ domains. When the P$_1$ domain is fully suppressed, the polarization should reach the value of $-P_s/3$$\approx$$-$0.8\,$\mu$C/cm$^2$. Thus, the smallest experimentally obtained value of +0.7\,$\mu$C/cm$^2$, recorded after poling with $-E_p$=7.4\,kV/cm, still points to a multi-domain state formed of all the four domain states. In parallel to the electric-field driven promotion (+$E_p$) or suppression ($-$$E_p$) of the P$_1$ domain, the two low-field peaks in the susceptibility are enhanced or suppressed, respectively, contrary to the two high-field anomalies.

Due to the coincidence of the magnetic easy axis with the polar axis~\cite{kezsmarki2015,ehlers2016}, the domain population can be efficiently controlled also by magnetic field. Poling with $H_p$ along the [111] axis is expected to promote the P$_1$ mono-domain state, irrespective of the sign of $H_p$, since the anisotropy energy is an even function of $H_p$. Indeed, Fig.~1e shows a continuous increase of polarization with increasing $H_p$ and its saturation for $\left|\mu_0H_p\right|$$\geq$9\,T, when the P$_1$ mono-domain state is achieved. The domain population, as probed via the magnetic susceptibility, shows the same trend in Fig.~1f.

Besides controlling and probing the volume fraction of the multiferroic domains, we investigated the influence of the domain population and DW density on the charge transport. The temperature-dependent resistivity in the mono-domain state, as obtained by poling with either +$E_p$ or $H_p$, follows a typical semiconducting behaviour both above and below the structural transition with only a tiny anomaly at $T_{JT}$ (see Fig.~2a). In contrast, the resistivity drops about four orders of magnitude at $T_{JT}$ in the multi-domain state, realized by poling with $-$$E_p$, and it is nearly independent of temperature below $T_{JT}$, implying the presence of delocalized charge carriers. Consequently, the low-temperature resistivity shows a huge difference between the mono- and multi-domain states, reaching eight orders of magnitude at $T$=15\,K. We attribute the enhanced conductivity of the multi-domain state to the presence of conductive DWs. With a smaller drop of the resistivity at $T_{JT}$, the unpoled crystal represents an intermediate case with a lower density of conductive DWs. This is in agreement with results in Fig.~1, which show the dominance of the P$_1$ domain over coexisting P$_2$, P$_3$, P$_4$ domains in the unpoled crystal.

\begin{figure}[t!]
\includegraphics[width=3.3in]{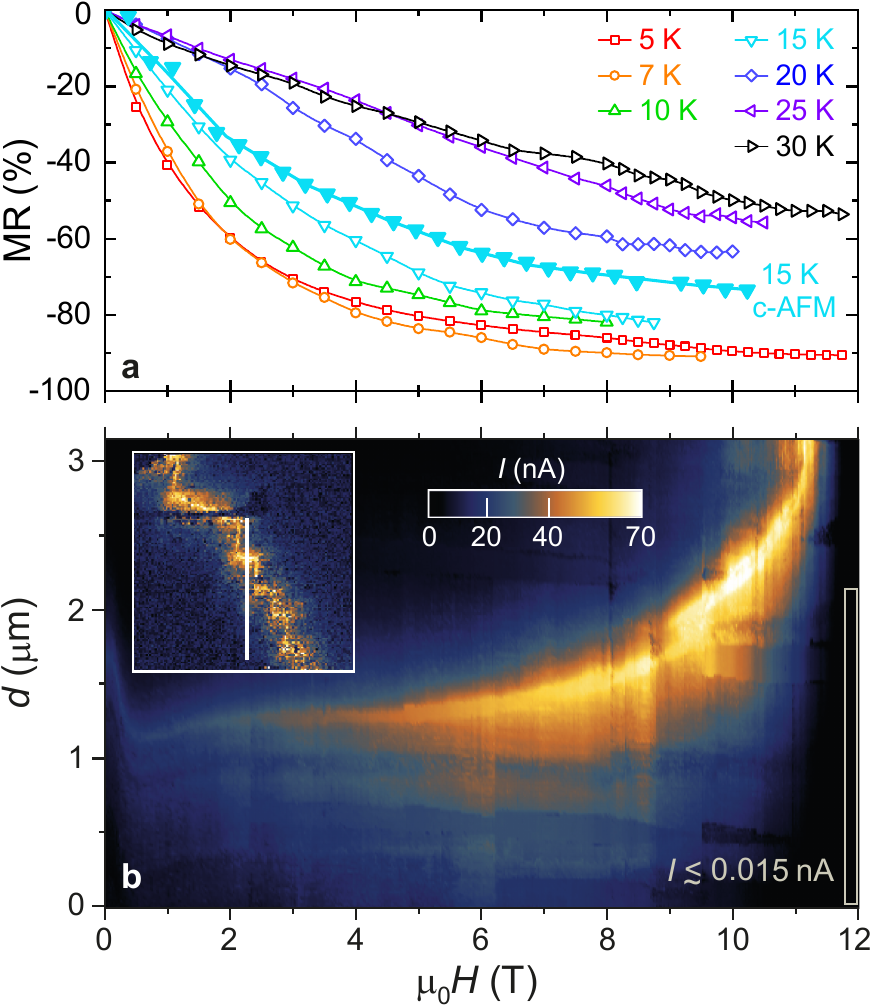}
\caption{\textbf{$\mid$ Giant negative MR of conductive DWs.} \textbf{a,} MR of a bulk multi-domain crystal at various temperatures, after poling with $E_p$ = $-$5.5 kV/cm. \textbf{b,} Image of c-AFM line-scan recorded while sweeping the magnetic field at 15\,K. The colour indicates the current for 45\,V applied between the tip and the sample. The white vertical line in the inset shows the location of the line-scan in a c-AFM image recorded in zero field. The typical current at the highest fields, the region within the grey frame, is $\lesssim$15\,pA. The MR deduced from the field-dependent c-AFM line-scans is also displayed in panel \textbf{a}.\label{fig3}}
\end{figure}

To identify the type of conductive DWs and to reveal the microscopic mechanism of DW conductivity, we carried out simultaneous PFM and c-AFM studies. Figures 2d and 2e respectively show a PFM and a c-AFM image, recorded at 15\,K on an as-grown (111) surface of an unpoled GaV$_4$S$_8$ crystal. Since the out-of-plane component of the piezoelectric effect is imaged, only the P$_1$ domain with polarization normal to this plane is distinguishable in Fig.~2d~\cite{butykai2017}. The other three domains, with polarizations all spanning 19$^{\circ}$ with the (111) plane, have the same out-of-plane piezoelectric signal. In the rhombohedral phase of GaV$_4$S$_8$, nine types of mechanically compatible DW can emerge: Three \{100\}-type uncharged and six \{110\}-type charged DWs (see supplementary Fig.~S1). On the (111) surface, we observed both \{100\}-type uncharged and \{110\}-type charged DWs, as shown in Fig.~2 and described in the following. The most obvious DW structure, seen on the left side of the PFM image, is a lamellar pattern of alternating brighter (P$_1$) and darker domains. To fulfill mechanical compatibility and charge neutrality, the dark ones must be P$_2$ domains, separated from P$_1$ domains by (010)-type head-to-tail (HT) DWs. The P$_2$ domain also forms a larger mono-domain island embedded in the P$_1$-P$_2$ lamellar pattern. On the right side of the PFM image, a P$_3$-P$_4$ lamellar pattern is seen with the same (010)-type HT DWs. Here, the two domains share the same colour, as expected, and only the DWs in between appear darker. Charge neutrality is confirmed for both lamellar patterns by the c-AFM image.

Next, we reveal fascinating architectures of the conductive DWs. The different domain patterns on the right and left side join in the middle to form a complex DW structure: The part below the green star separates a P$_2$ mono-domain island from the P$_3$-P$_4$ lamellar pattern, while the upper part separates the P$_1$-P$_2$ and the P$_3$-P$_4$ lamellar patterns, as also displayed schematically in Fig.~2c. The lower part shows up as lined-up bright spots in the c-AFM image, where the conductive spots are P$_2$-P$_3$ junctions, alternating with insulating P$_2$-P$_4$ segments. The P$_2$-P$_3$ junctions, being (1$\bar{1}$0)-type head-to-head (HH) conductive DWs are isolated, also electrically, by (100)-type HT DWs of the P$_2$-P$_4$ pairs. Thus, these short P$_2$-P$_3$ segments can appear as conductive spots in the c-AFM image only, if extending along the [001] axis as conductive ribbons. 

The upper part of the complex DW structure represents the typical 2D conductive DW sheets observed in this material (see supplementary Fig.~S2 for additional examples). It is composed of zig-zaging P$_2$-P$_3$ and P$_1$-P$_4$ junctions, where the former and the latter are (1$\bar{1}$0)-type HH and (110)-type tail-to-tail (TT) DWs, respectively.  Hence, this folded sheet, best resolved in Figs.~2f and 2g, is built of alternating stripes running along the [001] axis with very similar local conductivities but with opposite bound charges. Such a sharp alternation of TT and HH walls within a 2D folded conducting sheet is rather unique and differs from the situation in manganites, where these two types can be adjacent only in the vicinity of vortices\cite{choi2010,meier2012,wu2012}. (The notations TT, HT and HH refer to configurations of the normal component of the polarization across the 109$^{\circ}$ DWs.) The c-AFM data in Fig.~2e also confirm that the interior of the domains and the (010)-type DWs are insulating, though the latter can gain some conductance in proximity to charged DWs. For details about DW assignment, conductivity gain of (010)-type DWs and further PFM and c-AFM images see supplementary Figs.~S1-S2.

How do TT and HH DWs turn conductive in GaV$_4$S$_8$? Usually, charged DWs gain conductivity due to mobile carriers moving from the bulk to the DW region to screen the bound charge~\cite{vul1973,sluka2016}. In contrast to defects and off-stoichiometry supplying mobile carriers at the DWs of oxide ferroelectrics, the mobile holes/electrons, required to screen the negative/positive bound charge at TT/HH DWs, can be provided intrinsically in GaV$_4$S$_8$ due to the multivalent nature of V$_4$ clusters. The unique feature of the lacunar spinels is that the $B$-site cations constitute a molecular cluster~\cite{pocha2000,nakamura2005,dally2020}, which can readily accommodate different valency states. Indeed, the heterovalent substitution of Ga$^{3+}$ by Ge$^{4+}$ or by Zn$^{2+}$ has been demonstrated to respectively increase or decrease the number of $3d$ electrons occupying the V$_4$ molecular units~\cite{janod2015}. The stability of the two line compounds, GaV$_4$S$_8$ and GeV$_4$S$_8$, both categorized as narrow-gap molecular Mott insulators~\cite{reschke2017,reschke2020,kim2018}, also points to the high flexibility in the valency of the V$_4$ clusters. This intrinsic mechanism is sufficient to supply DWs with mobile carriers without requiring defects.

\begin{figure}[t!]
\includegraphics[width=3.1in]{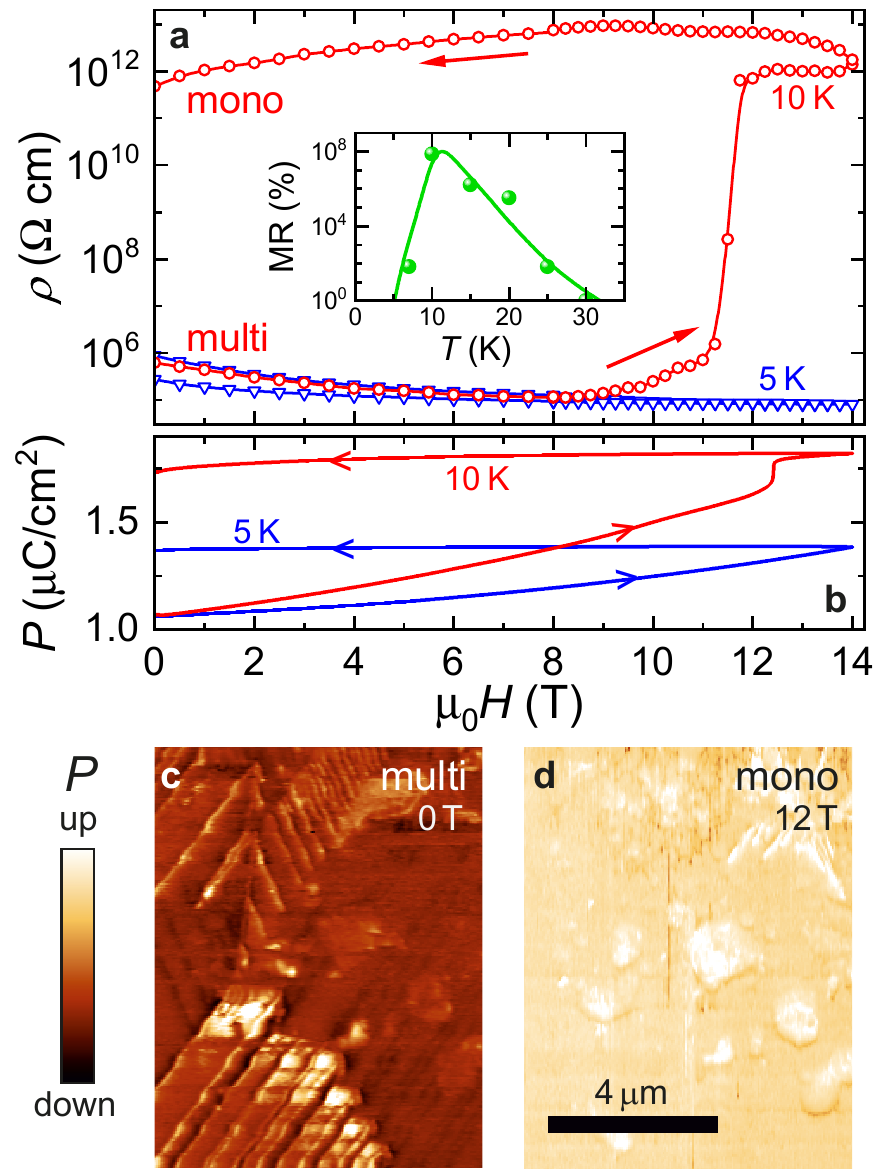}
\caption{\textbf{$\mid$ In situ magnetic erasure of conductive DWs.} \textbf{a,}
MR of a bulk multi-domain crystal at 5\,K and 10\,K, after poling with $E_p$ = $-$5.5 kV/cm. Irreversible magnetic switching from a low- to the high-resistance state was observed at 10\,K. The inset shows the difference of the zero-field resistance before and after the switching. \textbf{b,} Field-dependent polarization of a multi-domain crystal at 5\,K and 10\,K, after poling with $E_p$ = $-$5.5 kV/cm. At 10\,K it shows a jump at $\sim$12\,T, close to the field value where the resistive switching is observed. \textbf{c \& d}, PFM images recorded on the (111) surface of an unpoled crystal at 15\,K in a 0\,T multi-domain and the 12\,T mono-domain state, respectively.\label{fig4}}
\end{figure}

\textbf{Domain wall magnetotransport.} After characterizing the conductive DWs, we study the effect of magnetic field on their charge transport, which may be strong for multiferroic DWs~\cite{he2012,domingo2017,ma2015}. Figure 3a shows the negative MR of a multi-domain crystal at various temperatures, which gradually increases with decreasing temperature and reaches $\sim$90$\%$ in 12\,T at 5\,K. Since the resistivity in the multi-domain state is typically 4-8 orders of magnitude lower than in the mono-domain state, the conduction is fully dominated by the DWs and the giant negative MR is solely associated with the conducting DWs. This is directly confirmed by c-AFM line-scans measured while ramping the magnetic field. The tip-sample current, measured in subsequent line-scans across a conductive (110)-type DW and neutral (010)-type DWs adjacent to it, is shown in Fig.~3b as a function of the magnetic field. The current through the conductive DW is enhanced with increasing magnetic field, as indicated by the continuous brightening of the conducting region. The enhancement of the DW conductivity translates to a negative MR as high as $\sim$70$\%$ in 10\,T, which is fully consistent with the ``bulk" MR measured in multi-domain crystals (see Fig.~3a) and similar to that reported for conductive DWs in BiFeO$_3$~\cite{he2012}. In contrast, no conductivity enhancement was observed for either uncharged (010)-type DWs or the interior of the domains. In addition to its negative MR, the conductive DW is gradually displaced by the field and exits the scanned region above 11\,T, as seen in Fig.~3b. Only the magnitude of the conductivity and not the width of the conductive area around the DW is affected by the field, as becomes clear from supplementary Fig.~S3.

Does the magnetic field affect the density or the mobility of the carriers at DWs? Due to the multiferroic nature of the material, the polarization can be altered by magnetic field~\cite{ruff2015}. Since the field-driven polarization must be different for the different domain states, in principle, the charge density induced by the discontinuity of the polarization component normal to the wall could be tuned magnetically, which could charge originally neutral (010)-type DWs. For fields along the [111] axis, this effect should be the strongest for DWs involving the P$_1$ domain. However, we found that P$_1$-P$_2$ DWs remains insulating, excluding this scenario. Thus, we believe a likely explanation of the giant negative MR is the enhancement of carrier mobility due to the suppression of spin scattering by magnetic field. Since a field-polarized FM state, with magnetization co-aligned with the magnetic field and not with the easy axes of the domains, can be achieved in GaV$_4$S$_8$ by fields $\sim$1\,T, the magnetic state at conductive DWs should be distinct from that in the bulk, to leave room for spin-dependent scattering in fields $>$1\,T. Indeed, such magnetic states confined to DWs have been reported in GaV$_4$S$_8$~\cite{butykai2017} and GaV$_4$Se$_8$~\cite{geirhos2020}.

\textbf{Magnetically induced resistive switching.} Besides the giant MR of conductive DWs, magnetic fields $>$8\,T trigger irreversible changes, not shown in Fig.~3a. This phenomenon is best seen at 10\,K in the field-dependent resistivity and polarization curves respectively displayed in Figs.~4a \& 4b. (For data at various temperatures see supplementary Fig.~S4). The resistivity shows a six orders of magnitude step-like increase at around 12\,T, accompanied with a polarization jump at slightly higher fields. The sudden jumps in the two quantities, measured in subsequent poling runs, indicate the in-situ magnetic switching from a low-resistance multi-domain to the high-resistance mono-domain state. In Fig.~3b, the highly accelerated movement of the DW above 10\,T  also hints at a catastrophic DW expulsion event, eventually leading to the mono-domain state. In fact, the low current values ($\lesssim$15\,pA) detected in line scans at the highest fields roughly correspond to the typical resistance found in macroscopic transport measurements on mono-domain crystals. Moreover, at 15\,K the ratio of the resistance values with the tip placed on a highly conductive DW in an unpoled sample and on the surface of a mono-domain crystal is approximately 10$^4$, which is close to the ratio observed in macroscopic measurements at the same temperature in Fig.~2a.

PFM data in Figs.~4c \& 4d directly support this scenario: The multi-domain pattern, observed in the unpoled crystal, is turned to a P$_1$ mono-domain in $\mu_0H$ $\approx$12\,T, which persists after switching off the field. The inset of Fig.~4a displays the magnitude of the resistivity jump versus temperature, measuring the efficiency of the in-situ magnetic switching. It shows a dramatic increase upon approaching the magnetic ordering temperature, followed by a fast drop due to the reduced mobility of the DWs at the lowest temperatures. In fact, no field-induced switching takes place at 5\,K, though the polarization gradually increases up to 14\,T and stays at the highest value when reducing the field to zero. This still implies some reconfigurations of the domain structure, yet the density of conductive DWs is not reduced below the percolation limit.

\textbf{Conclusions.} In summary, we demonstrated the fascinating properties and in-situ electric and magnetic control of highly conductive DWs in a non-oxide material, the multiferroic GaV$_4$S$_8$. These DWs can host either conduction electrons or holes, both of which are intrinsically provided by the multivalent V$_4$ molecular clusters. As a unique attribute of these DWs, they fully dominate the electrical transport of the material. Consequently, their efficient manipulation enables on-demand gigantic switching of the sample resistance by several orders of magnitude. In addition, their versatile architectures provide new routes to design and control complex networks of conductive DWs. For example, the immediate vicinity of electron- and hole-like conductive segments is not only fascinating, but also allows for the first time to design $p$ and $n$-type elements in nanoscale proximity and lets us envision $p-n$-junction-like functional objects in non-oxide conducting DWs. These findings shall trigger an extensive search for novel non-oxide materials hosting easy-to-control topological defects and exotic electronic and magnetic states associated with them.

\bibliography{GaV4S8_conductive_DWs_arxiv}

\begin{thebibliography}{10}
\expandafter\ifx\csname url\endcsname\relax
  \def\url#1{\texttt{#1}}\fi
\expandafter\ifx\csname urlprefix\endcsname\relax\def\urlprefix{URL }\fi
\providecommand{\bibinfo}[2]{#2}
\providecommand{\eprint}[2][]{\url{#2}}

\bibitem{catalan2012}
\bibinfo{author}{Catalan, G.}, \bibinfo{author}{Seidel, J.},
  \bibinfo{author}{Ramesh, R.} \& \bibinfo{author}{Scott, J.~F.}
\newblock \bibinfo{title}{Domain wall nanoelectronics}.
\newblock \emph{\bibinfo{journal}{Rev. Mod. Phys.}}
  \textbf{\bibinfo{volume}{84}}, \bibinfo{pages}{119} (\bibinfo{year}{2012}).

\bibitem{erhart2004}
\bibinfo{author}{Erhart, J.}
\newblock \bibinfo{title}{Domain wall orientations in ferroelastics and
  ferroelectrics}.
\newblock \emph{\bibinfo{journal}{Phase Transitions}}
  \textbf{\bibinfo{volume}{77}}, \bibinfo{pages}{989--1074}
  (\bibinfo{year}{2004}).

\bibitem{aird1998}
\bibinfo{author}{Aird, A.} \& \bibinfo{author}{Salje, E.~K.}
\newblock \bibinfo{title}{Sheet superconductivity in twin walls: experimental
  evidence of {WO$_{3-x}$}}.
\newblock \emph{\bibinfo{journal}{J. Phys. Condens. Matter}}
  \textbf{\bibinfo{volume}{10}}, \bibinfo{pages}{L377} (\bibinfo{year}{1998}).

\bibitem{salje2010}
\bibinfo{author}{Salje, E.~K.}
\newblock \bibinfo{title}{Multiferroic domain boundaries as active memory
  devices: trajectories towards domain boundary engineering}.
\newblock \emph{\bibinfo{journal}{ChemPhysChem}} \textbf{\bibinfo{volume}{11}},
  \bibinfo{pages}{940--950} (\bibinfo{year}{2010}).

\bibitem{farokhipoor2014}
\bibinfo{author}{Farokhipoor, S.} \emph{et~al.}
\newblock \bibinfo{title}{Artificial chemical and magnetic structure at the
  domain walls of an epitaxial oxide}.
\newblock \emph{\bibinfo{journal}{Nature}} \textbf{\bibinfo{volume}{515}},
  \bibinfo{pages}{379--383} (\bibinfo{year}{2014}).

\bibitem{van2012}
\bibinfo{author}{Van~Aert, S.} \emph{et~al.}
\newblock \bibinfo{title}{Direct observation of ferrielectricity at
  ferroelastic domain boundaries in {CaTiO$_3$} by electron microscopy}.
\newblock \emph{\bibinfo{journal}{Adv. Mater.}} \textbf{\bibinfo{volume}{24}},
  \bibinfo{pages}{523--527} (\bibinfo{year}{2012}).

\bibitem{zubko2007}
\bibinfo{author}{Zubko, P.}, \bibinfo{author}{Catalan, G.},
  \bibinfo{author}{Buckley, A.}, \bibinfo{author}{Welche, P.} \&
  \bibinfo{author}{Scott, J.}
\newblock \bibinfo{title}{Strain-gradient-induced polarization in {SrTiO$_3$}
  single crystals}.
\newblock \emph{\bibinfo{journal}{Phys. Rev. Lett.}}
  \textbf{\bibinfo{volume}{99}}, \bibinfo{pages}{167601}
  (\bibinfo{year}{2007}).

\bibitem{vul1973}
\bibinfo{author}{Vul, B.}, \bibinfo{author}{Guro, G.} \&
  \bibinfo{author}{Ivanchik, I.}
\newblock \bibinfo{title}{Encountering domains in ferroelectrics}.
\newblock \emph{\bibinfo{journal}{Ferroelectrics}}
  \textbf{\bibinfo{volume}{6}}, \bibinfo{pages}{29--31} (\bibinfo{year}{1973}).

\bibitem{gureev2011}
\bibinfo{author}{Gureev, M.~Y.}, \bibinfo{author}{Tagantsev, A.~K.} \&
  \bibinfo{author}{Setter, N.}
\newblock \bibinfo{title}{Head-to-head and tail-to-tail 180{$^{\circ}$} domain
  walls in an isolated ferroelectric}.
\newblock \emph{\bibinfo{journal}{Phys. Rev. B}} \textbf{\bibinfo{volume}{83}},
  \bibinfo{pages}{184104} (\bibinfo{year}{2011}).

\bibitem{eliseev2011}
\bibinfo{author}{Eliseev, E.}, \bibinfo{author}{Morozovska, A.},
  \bibinfo{author}{Svechnikov, G.}, \bibinfo{author}{Gopalan, V.} \&
  \bibinfo{author}{Shur, V.~Y.}
\newblock \bibinfo{title}{Static conductivity of charged domain walls in
  uniaxial ferroelectric semiconductors}.
\newblock \emph{\bibinfo{journal}{Phys. Rev. B}} \textbf{\bibinfo{volume}{83}},
  \bibinfo{pages}{235313} (\bibinfo{year}{2011}).

\bibitem{meier2015}
\bibinfo{author}{Meier, D.}
\newblock \bibinfo{title}{Functional domain walls in multiferroics}.
\newblock \emph{\bibinfo{journal}{J. Phys. Condens. Matter}}
  \textbf{\bibinfo{volume}{27}}, \bibinfo{pages}{463003}
  (\bibinfo{year}{2015}).

\bibitem{sluka2016}
\bibinfo{author}{Sluka, T.}, \bibinfo{author}{Bednyakov, P.},
  \bibinfo{author}{Yudin, P.}, \bibinfo{author}{Crassous, A.} \&
  \bibinfo{author}{Tagantsev, A.}
\newblock \bibinfo{title}{Charged domain walls in ferroelectrics}.
\newblock In \emph{\bibinfo{booktitle}{Topological structures in ferroic
  materials}}, \bibinfo{pages}{103--138} (\bibinfo{publisher}{Springer},
  \bibinfo{year}{2016}).

\bibitem{sluka2013}
\bibinfo{author}{Sluka, T.}, \bibinfo{author}{Tagantsev, A.~K.},
  \bibinfo{author}{Bednyakov, P.} \& \bibinfo{author}{Setter, N.}
\newblock \bibinfo{title}{Free-electron gas at charged domain walls in
  insulating {BaTiO$_3$}}.
\newblock \emph{\bibinfo{journal}{Nat. Commun.}} \textbf{\bibinfo{volume}{4}},
  \bibinfo{pages}{1--6} (\bibinfo{year}{2013}).

\bibitem{schroder2012}
\bibinfo{author}{Schr{\"o}der, M.} \emph{et~al.}
\newblock \bibinfo{title}{Conducting domain walls in lithium niobate single
  crystals}.
\newblock \emph{\bibinfo{journal}{Adv. Funct. Mater.}}
  \textbf{\bibinfo{volume}{22}}, \bibinfo{pages}{3936--3944}
  (\bibinfo{year}{2012}).

\bibitem{werner2017l}
\bibinfo{author}{Werner, C.~S.} \emph{et~al.}
\newblock \bibinfo{title}{Large and accessible conductivity of charged domain
  walls in lithium niobate}.
\newblock \emph{\bibinfo{journal}{Sci. Rep.}} \textbf{\bibinfo{volume}{7}},
  \bibinfo{pages}{1--8} (\bibinfo{year}{2017}).

\bibitem{meier2012}
\bibinfo{author}{Meier, D.} \emph{et~al.}
\newblock \bibinfo{title}{Anisotropic conductance at improper ferroelectric
  domain walls}.
\newblock \emph{\bibinfo{journal}{Nat. Mater.}} \textbf{\bibinfo{volume}{11}},
  \bibinfo{pages}{284--288} (\bibinfo{year}{2012}).

\bibitem{wu2012}
\bibinfo{author}{Wu, W.}, \bibinfo{author}{Horibe, Y.}, \bibinfo{author}{Lee,
  N.}, \bibinfo{author}{Cheong, S.-W.} \& \bibinfo{author}{Guest, J.}
\newblock \bibinfo{title}{Conduction of topologically protected charged
  ferroelectric domain walls}.
\newblock \emph{\bibinfo{journal}{Phys. Rev. Lett.}}
  \textbf{\bibinfo{volume}{108}}, \bibinfo{pages}{077203}
  (\bibinfo{year}{2012}).

\bibitem{seidel2009}
\bibinfo{author}{Seidel, J.} \emph{et~al.}
\newblock \bibinfo{title}{Conduction at domain walls in oxide multiferroics}.
\newblock \emph{\bibinfo{journal}{Nat. Mater.}} \textbf{\bibinfo{volume}{8}},
  \bibinfo{pages}{229--234} (\bibinfo{year}{2009}).

\bibitem{farokhipoor2011}
\bibinfo{author}{Farokhipoor, S.} \& \bibinfo{author}{Noheda, B.}
\newblock \bibinfo{title}{Conduction through {71$^{\circ}$} domain walls in
  {BiFeO$_3$} thin films}.
\newblock \emph{\bibinfo{journal}{Phys. Rev. Lett.}}
  \textbf{\bibinfo{volume}{107}}, \bibinfo{pages}{127601}
  (\bibinfo{year}{2011}).

\bibitem{johrendt1998}
\bibinfo{author}{Johrendt, D.}
\newblock \bibinfo{title}{Crystal and electronic structure of the tetrahedral
  v4 cluster compounds {GeV$_4$Q$_8$} ({Q}= {S}, {Se})}.
\newblock \emph{\bibinfo{journal}{Z. Anorg. Allg. Chem.}}
  \textbf{\bibinfo{volume}{624}}, \bibinfo{pages}{952--958}
  (\bibinfo{year}{1998}).

\bibitem{pocha2000}
\bibinfo{author}{Pocha, R.}, \bibinfo{author}{Johrendt, D.} \&
  \bibinfo{author}{P{\"o}ttgen, R.}
\newblock \bibinfo{title}{Electronic and structural instabilities in
  {GaV$_4$S$_8$} and {GaMo$_4$S$_8$}}.
\newblock \emph{\bibinfo{journal}{Chem. Mater.}} \textbf{\bibinfo{volume}{12}},
  \bibinfo{pages}{2882--2887} (\bibinfo{year}{2000}).

\bibitem{nakamura2005}
\bibinfo{author}{Nakamura, H.}, \bibinfo{author}{Chudo, H.} \&
  \bibinfo{author}{Shiga, M.}
\newblock \bibinfo{title}{Structural transition of the tetrahedral metal
  cluster: nuclear magnetic resonance study of {GaV$_4$S$_8$}}.
\newblock \emph{\bibinfo{journal}{J. Phys. Condens. Matter}}
  \textbf{\bibinfo{volume}{17}}, \bibinfo{pages}{6015} (\bibinfo{year}{2005}).

\bibitem{muller2006}
\bibinfo{author}{M{\"u}ller, H.}, \bibinfo{author}{Kockelmann, W.} \&
  \bibinfo{author}{Johrendt, D.}
\newblock \bibinfo{title}{The magnetic structure and electronic ground states
  of mott insulators {GeV$_4$S$_8$} and {GaV$_4$S$_8$}}.
\newblock \emph{\bibinfo{journal}{Chem. Mater.}} \textbf{\bibinfo{volume}{18}},
  \bibinfo{pages}{2174--2180} (\bibinfo{year}{2006}).

\bibitem{singh2014}
\bibinfo{author}{Singh, K.} \emph{et~al.}
\newblock \bibinfo{title}{Orbital-ordering-driven multiferroicity and
  magnetoelectric coupling in {GeV$_4$S$_8$}}.
\newblock \emph{\bibinfo{journal}{Phys. Rev. Lett.}}
  \textbf{\bibinfo{volume}{113}}, \bibinfo{pages}{137602}
  (\bibinfo{year}{2014}).

\bibitem{janod2015}
\bibinfo{author}{Janod, E.} \emph{et~al.}
\newblock \bibinfo{title}{Negative colossal magnetoresistance driven by carrier
  type in the ferromagnetic mott insulator {GaV$_4$S$_8$}}.
\newblock \emph{\bibinfo{journal}{Chem. Mater.}} \textbf{\bibinfo{volume}{27}},
  \bibinfo{pages}{4398--4404} (\bibinfo{year}{2015}).

\bibitem{reschke2020}
\bibinfo{author}{Reschke, S.} \emph{et~al.}
\newblock \bibinfo{title}{Lattice dynamics and electronic excitations in a
  large family of lacunar spinels with a breathing pyrochlore lattice
  structure}.
\newblock \emph{\bibinfo{journal}{Phys. Rev. B}}
  \textbf{\bibinfo{volume}{101}}, \bibinfo{pages}{075118}
  (\bibinfo{year}{2020}).

\bibitem{butykai2017}
\bibinfo{author}{Butykai, {\'A}.} \emph{et~al.}
\newblock \bibinfo{title}{Characteristics of ferroelectric-ferroelastic domains
  in {N{\'e}el}-type skyrmion host {GaV$_4$S$_8$}}.
\newblock \emph{\bibinfo{journal}{Sci. Rep.}} \textbf{\bibinfo{volume}{7}},
  \bibinfo{pages}{44663} (\bibinfo{year}{2017}).

\bibitem{neuber2018}
\bibinfo{author}{Neuber, E.} \emph{et~al.}
\newblock \bibinfo{title}{Architecture of nanoscale ferroelectric domains in
  {GaMo$_4$S$_8$}}.
\newblock \emph{\bibinfo{journal}{J. Phys. Condens. Matter}}
  \textbf{\bibinfo{volume}{30}}, \bibinfo{pages}{445402}
  (\bibinfo{year}{2018}).

\bibitem{yadav2008}
\bibinfo{author}{Yadav, C.}, \bibinfo{author}{Nigam, A.} \&
  \bibinfo{author}{Rastogi, A.}
\newblock \bibinfo{title}{Thermodynamic properties of ferromagnetic
  mott-insulator {GaV$_4$S$_8$}}.
\newblock \emph{\bibinfo{journal}{Physica B: Condensed Matter}}
  \textbf{\bibinfo{volume}{403}}, \bibinfo{pages}{1474--1475}
  (\bibinfo{year}{2008}).

\bibitem{nakamura2009}
\bibinfo{author}{Nakamura, H.} \emph{et~al.}
\newblock \bibinfo{title}{Low-field multi-step magnetization of {GaV$_4$S$_8$}
  single crystal}.
\newblock \emph{\bibinfo{journal}{J. Phys. Conf. Ser.}}
  \textbf{\bibinfo{volume}{145}}, \bibinfo{pages}{012077}
  (\bibinfo{year}{2009}).

\bibitem{kezsmarki2015}
\bibinfo{author}{K{\'e}zsm{\'a}rki, I.} \emph{et~al.}
\newblock \bibinfo{title}{{N{\'e}el}-type skyrmion lattice with confined
  orientation in the polar magnetic semiconductor {GaV$_4$S$_8$}}.
\newblock \emph{\bibinfo{journal}{Nat. Mater.}} \textbf{\bibinfo{volume}{14}},
  \bibinfo{pages}{1116--1122} (\bibinfo{year}{2015}).

\bibitem{ruff2015}
\bibinfo{author}{Ruff, E.} \emph{et~al.}
\newblock \bibinfo{title}{Multiferroicity and skyrmions carrying electric
  polarization in {GaV$_4$S$_8$}}.
\newblock \emph{\bibinfo{journal}{Sci. Adv.}} \textbf{\bibinfo{volume}{1}},
  \bibinfo{pages}{e1500916} (\bibinfo{year}{2015}).

\bibitem{widmann2017}
\bibinfo{author}{Widmann, S.} \emph{et~al.}
\newblock \bibinfo{title}{On the multiferroic skyrmion-host {GaV$_4$S$_8$}}.
\newblock \emph{\bibinfo{journal}{Philos. Mag.}} \textbf{\bibinfo{volume}{97}},
  \bibinfo{pages}{3428--3445} (\bibinfo{year}{2017}).

\bibitem{ehlers2016}
\bibinfo{author}{Ehlers, D.} \emph{et~al.}
\newblock \bibinfo{title}{Exchange anisotropy in the skyrmion host
  {GaV$_4$S$_8$}}.
\newblock \emph{\bibinfo{journal}{J. Phys. Condens. Matter}}
  \textbf{\bibinfo{volume}{29}}, \bibinfo{pages}{065803}
  (\bibinfo{year}{2016}).

\bibitem{choi2010}
\bibinfo{author}{Choi, T.} \emph{et~al.}
\newblock \bibinfo{title}{Insulating interlocked ferroelectric and structural
  antiphase domain walls in multiferroic {YMnO$_3$}}.
\newblock \emph{\bibinfo{journal}{Nat. Mater.}} \textbf{\bibinfo{volume}{9}},
  \bibinfo{pages}{253--258} (\bibinfo{year}{2010}).

\bibitem{dally2020}
\bibinfo{author}{Dally, R.~L.} \emph{et~al.}
\newblock \bibinfo{title}{Magnetic phase transitions and spin density
  distribution in the molecular multiferroic system {GaV$_4$S$_8$}}.
\newblock \emph{\bibinfo{journal}{Phys. Rev. B}}
  \textbf{\bibinfo{volume}{102}}, \bibinfo{pages}{014410}
  (\bibinfo{year}{2020}).

\bibitem{reschke2017}
\bibinfo{author}{Reschke, S.} \emph{et~al.}
\newblock \bibinfo{title}{Optical conductivity in multiferroic {GaV$_4$S$_8$}
  and {GeV$_4$S$_8$}: Phonons and electronic transitions}.
\newblock \emph{\bibinfo{journal}{Phys. Rev. B}} \textbf{\bibinfo{volume}{96}},
  \bibinfo{pages}{144302} (\bibinfo{year}{2017}).

\bibitem{kim2018}
\bibinfo{author}{Kim, H.-S.}, \bibinfo{author}{Haule, K.} \&
  \bibinfo{author}{Vanderbilt, D.}
\newblock \bibinfo{title}{Molecular {Mott} state in the deficient spinel
  {GaV$_4$S$_8$}}.
\newblock \emph{\bibinfo{journal}{arXiv preprint arXiv:1810.09495}}
  (\bibinfo{year}{2018}).

\bibitem{he2012}
\bibinfo{author}{He, Q.} \emph{et~al.}
\newblock \bibinfo{title}{Magnetotransport at domain walls in {BiFeO$_3$}}.
\newblock \emph{\bibinfo{journal}{Phys. Rev. Lett.}}
  \textbf{\bibinfo{volume}{108}}, \bibinfo{pages}{067203}
  (\bibinfo{year}{2012}).

\bibitem{domingo2017}
\bibinfo{author}{Domingo, N.}, \bibinfo{author}{Farokhipoor, S.},
  \bibinfo{author}{Santiso, J.}, \bibinfo{author}{Noheda, B.} \&
  \bibinfo{author}{Catalan, G.}
\newblock \bibinfo{title}{Domain wall magnetoresistance in {BiFeO$_3$} thin
  films measured by scanning probe microscopy}.
\newblock \emph{\bibinfo{journal}{J. Phys. Condens. Matter}}
  \textbf{\bibinfo{volume}{29}}, \bibinfo{pages}{334003}
  (\bibinfo{year}{2017}).

\bibitem{ma2015}
\bibinfo{author}{Ma, E.~Y.} \emph{et~al.}
\newblock \bibinfo{title}{Mobile metallic domain walls in an all-in-all-out
  magnetic insulator}.
\newblock \emph{\bibinfo{journal}{Science}} \textbf{\bibinfo{volume}{350}},
  \bibinfo{pages}{538--541} (\bibinfo{year}{2015}).

\bibitem{geirhos2020}
\bibinfo{author}{Geirhos, K.} \emph{et~al.}
\newblock \bibinfo{title}{Macroscopic manifestation of domain-wall magnetism
  and magnetoelectric effect in a {N{\'e}el}-type skyrmion host}.
\newblock \emph{\bibinfo{journal}{npj Quantum Materials}}
  \textbf{\bibinfo{volume}{5}}, \bibinfo{pages}{1--8} (\bibinfo{year}{2020}).

\end{thebibliography}

\section*{Methods}

\textbf{Polarization measurements.}~All experiments were performed on insulating single crystals of GaV$_4$S$_8$, which were prepared by the chemical vapour transport method, using iodine as a transport agent~\cite{kezsmarki2015}. Electric polarization was recorded by conventional pyroelectric current measurements using a Keysight electrometer. For this purpose, a (111)-cut crystal was contacted by silver paste in a top-bottom capacitor geometry. The poling fields, either electric or magnetic, were applied along [111] direction at 50\,K and the sample was cooled to 4\,K, where the poling fields were switched off and the sample was kept shorted for 15\,minutes. Pyroelectric current was measured during heating with a heating-rate of 7\,K/min. Magnetic field was swept at a rate of 1\,T/min when recording the magneto-current in the magnetic field dependent polarization studies. The polarization was obtained by integrating the pyroelectric current or magneto-current over time.

\textbf{Resistivity measurements.}~The dc resistivity was measured by the four-probe method, where a Keithley electrometer (6517B) was used to apply a source voltage of $+/-$8\,V between terminal 1 and 2, located at opposite (111) faces of the sample, as well as to measure current through these terminals. The voltage drop between terminals 3 and 4 was recorded by a Keysight electrometer (B2987A). Terminal 3 and 4 was adjacent to and co-planar with terminal 1 and 2, respectively. Each co-planar contact pair nearly covered the corresponding surface with a small gap between them. During the poling procedures, the terminals, 1 and 3 as well as terminals 2 and 4, were shorted and the poling fields were applied from 50\,K down to 4\,K , where the poling fields were switched off before starting the measurements during heating the sample. For temperature- and field-dependent measurements, resistivity was recorded after stabilizing the temperature and the magnetic field in each step in order to avoid pyroelectric and magnetoelectric contributions, respectively.

\textbf{Magnetic measurements.}~The dc magnetization measurements were carried out with a Physical Property Measurement System (14\,T - PPMS) of Quantum Design, while the ac susceptibility was measured with a Magnetic Property Measurement System (5\,T MPMS - SQUID) of Quantum Design. For electric-field poling, a home-designed probe was developed for the MPMS system and the same poling protocol was followed as for the polarization measurements.

\textbf{PFM and c-AFM.}~Cryogenic AFM measurements were performed in an attoLiquid2000 setup in an atmosphere of 20\,mbar He exchange gas using conductive Pt/Ir-coated ANSCM-Pt tips. The unique cryogenic setup allows for scanning probe studies in magnetic fields up to 12\,T, but only supports the detection of the out-of-plane piezoelectric coefficient. Measurements were
performed on an as-grown (111)-face of the crystal, with a polished parallel (111)-face of the sample attached to a back electrode. For PFM measurements, an external Stanford Instruments SR830 lock-in amplifier was used to apply an excitation voltage of 5\,V to the tip at a fixed frequency of 17.87\,kHz, while the back electrode was grounded. The signal was optimized for maximum domain contrast by setting the phase shift such that all signal appears in the X-channel, when scanning the interior of the domains. The c-AFM measurements were performed by applying dc bias voltages of up to 60\,V to the bottom electrode and detecting the current flowing through the tip with a FEMTO I/V converter.

\textbf{Acknowledgements}~We thank Anton Jesche and Dana Vieweg for assistance in the magnetization measurements and S\'andor Bord\'acs, \'Ad\'am Butykai, Stephan Krohns, Donald Evans, Lukas Puntigam, Peter Milde, Lukas Eng and Jiri Hlinka for fruitful discussions. This work was supported by the DFG via the Transregional Research Collaboration TRR 80 From Electronic Correlations to Functionality (Augsburg/Munich/Stuttgart) and via the DFG Priority Program SPP2137, Skyrmionics, under Grant No. KE 2370/1-1 and by the Swiss National Science Foundation under Grant Nos. SNSF 206021\_150635 and 200021\_178825. V.T. acknowledges the project ANCD 20.80009.5007.19 (Moldova).

\textbf{Author Contributions:}~S.G., K.G. and P.L. performed and analyzed the polarization and resistivity measurements; S.G. performed the magnetic measurements; V.T. synthesized the single crystals; L.K. and M.F. performed PFM and c-AFM experiments and the data were analysed together with S.G. and K.G.; S.G. and I.K. wrote the manuscript with contributions from L.K., K.G. and M.F.; I.K. supervised this project. All authors discussed the content of the manuscript. 

\textbf{Additional information:}~The authors declare no competing financial interests.

\end{document}